\documentclass[twocolumn,superscriptaddress,aps,prl,amsmath,amssymb,floatfix]{revtex4-1}
\usepackage{amsmath,amsthm,amssymb}
\usepackage{latexsym}
\usepackage{graphicx}
\usepackage[normalem]{ulem}
\usepackage[dvipsnames]{xcolor}
\usepackage[colorlinks=true,linkcolor=red,citecolor=red,urlcolor=red]{hyperref}
\usepackage{physics}
\usepackage{enumerate}

\begin{document}

\title{Strong coupling between {a single-photon and a two-photon Fock state}}

\author{Shuai-Peng Wang}
\affiliation{Beijing Key Laboratory of Fault-Tolerant Quantum Computing, Beijing Academy of Quantum Information Sciences, Beijing 100193, China}
\affiliation{Quantum Physics and Quantum Information Division, Beijing Computational Science Research Center, Beijing 100193, China}
\affiliation{Zhejiang Key Laboratory of Micro-Nano Quantum Chips and Quantum Control, School of Physics, Zhejiang University, Hangzhou 310027, China}

\author{Alberto Mercurio}
\affiliation{Institute of Physics, {\'E}cole Polytechnique F{\'e}d{\'e}rale de Lausanne (EPFL), CH-1015 Lausanne, Switzerland}
\affiliation{Center for Quantum Science and Engineering, {\'E}cole Polytechnique F{\'e}d{\'e}rale de Lausanne (EPFL), CH-1015 Lausanne, Switzerland}

\author{Alessandro Ridolfo}
\affiliation{Dipartimento di Fisica e Astronomia, Universit{\`a} di Catania, 95123 Catania, Italy}

\author{Yuqing Wang}
\affiliation{Beijing Key Laboratory of Fault-Tolerant Quantum Computing, Beijing Academy of Quantum Information Sciences, Beijing 100193, China}

\author{Mo Chen}
\affiliation{Beijing Key Laboratory of Fault-Tolerant Quantum Computing, Beijing Academy of Quantum Information Sciences, Beijing 100193, China}

\author{Wenyan Wang}
\affiliation{Beijing Key Laboratory of Fault-Tolerant Quantum Computing, Beijing Academy of Quantum Information Sciences, Beijing 100193, China}

\author{Yulong Liu}
\affiliation{Beijing Key Laboratory of Fault-Tolerant Quantum Computing, Beijing Academy of Quantum Information Sciences, Beijing 100193, China}

\author{Huanying Sun}
\affiliation{Beijing Key Laboratory of Fault-Tolerant Quantum Computing, Beijing Academy of Quantum Information Sciences, Beijing 100193, China}

\author{Tiefu Li}
\email{litf@tsinghua.edu.cn}
\affiliation{School of Integrated Circuits, and Frontier Science Center for Quantum Information, Tsinghua University, Beijing 100084, China}
%\affiliation{Beijing Academy of Quantum Information Sciences, Beijing 100193, China}

\author{Franco Nori}
\affiliation{Quantum Computing Center, RIKEN, Wako-shi, Saitama 351-0198, Japan}
\affiliation{Physics Department, The University of Michigan, Ann Arbor, Michigan 48109-1040, USA}

\author{Salvatore Savasta}
\email{salvatore.savasta@unime.it}
\affiliation{Dipartimento di Scienze Matematiche e Informatiche, Scienze Fisiche e Scienze della Terra, Universit{\`a} di Messina, I-98166 Messina, Italy}

\author{J. Q. You}
\email{jqyou@zju.edu.cn}
\affiliation{Zhejiang Key Laboratory of Micro-Nano Quantum Chips and Quantum Control, School of Physics, Zhejiang University, Hangzhou 310027, China}

\begin{abstract}

\noindent
The realization of strong nonlinear coupling between single photons has been a long-standing goal in quantum optics and quantum information science, promising wide impact applications, such as all-optical deterministic quantum logic and single-photon frequency conversion. Here, we report an experimental observation of the strong coupling between a single-photon and a two-photon Fock state in an ultrastrongly-coupled circuit-QED system. This strong nonlinear interaction is realized by introducing a detuned flux qubit working as an effective coupler between two modes of a superconducting coplanar waveguide resonator. The ultrastrong light--matter interaction breaks the excitation number conservation, and an external flux bias breaks the parity conservation. The combined effect of the two enables the strong one--two-photon coupling. Quantum Rabi-like avoided crossing is resolved when tuning the two-photon resonance frequency of the first mode across the single-photon resonance frequency of the second mode. Within this new photonic regime, we observe the thresholdless second harmonic generation for a mean photon number below one. Our results represent a key step towards a new regime of quantum nonlinear optics, where individual photons can  deterministically and coherently interact with each other in the absence of any stimulating fields.
\end{abstract}

%\date{\today}

\maketitle

\noindent{\bf \large Introduction}\\
Strong interactions between single photons enable numerous applications in quantum information processing and simulations~\cite{Lukin2014NPh}. While photons in a vacuum do not interact with each other, when they are confined in nonlinear optical media, the light--matter interaction results in an effective photon--photon interaction~\cite{Scully1999, Agarwal2012}. However, the nonlinearity of conventional materials is negligibly weak at the single-photon power level, making realizing strong nonlinear coupling between single photons extremely hard~\cite{Lukin2014NPh}.

So far, important nonlinear optical phenomena like second harmonic generation (SHG) or down-conversion (DC) occur only statistically and in the presence of many photons, so a more or less small fraction of photons undergo fusion or fission process. Taking advantage of the resonant interaction between cavity photons and atoms in cavity QED~\cite{Kimble2005Nature}, using electromagnetically induced transparency (EIT)~\cite{Lukin2009PRL}, or collective effects in atomic ensembles~\cite{Vuletic2012Nature}, and applying high-power stimulating fields~\cite{Aumentado2011NP} are the common ways to circumvent this difficulty and to bring  photon--photon interactions close to the quantum limit. However, an experimental observation of the strong coupling between different photon-number states~\cite{Lukin2014NPh, Bouwmeester2006PRL} in the absence of control fields with many photons, a cornerstone of quantum many-body nonlinear optics, is still elusive.

The development of superconducting quantum circuits (e.g.,~\cite{Martinis1985PRL, Nakamura1999Nature, Mooij1999Science, Martinis2002PRL, You2007PRB, Koch2007PRA}) brings in new possibilities to tackle this problem. Josephson-junction (JJ)-based artificial atoms have a large dipole moment (not limited by the fine structure constant), which, combined with the small mode volume of superconducting coplanar resonators, allows the very large photon--artificial-atom interactions~\cite{You2011Nature, Nori2017PhyRep, Mooij2004Nature, Schoelkopf2004Nature, Oliver2020ARCMP}. The largest reported single-photon--single-qubit interactions have entered the ultrastrong and even deep strong coupling regimes~\cite{Gross2010NP, Mooij2010PRL, Lupascu2017NP, Semba2017NP, Semba2018PRL, Chen2017PRA, Wang2020PRApplied, Wang2023NC, Tomonaga2023arXiv, Nori2019NRP, Solano2019RMP, Hanggi2009PRA, Nori2010PRA, Solano2010PRL}. This remarkable property makes Josephson artificial atoms an ideal effective medium to realize strong nonlinear coupling between single photons~\cite{Nori2017PRA, Semba2022PRRe, Brien2021PRL, Schoelkopf2013Nature, Manucharyan2022Nature, Fraudet2024arXiv}.

%==============================================
\begin{figure*}[t]
\includegraphics{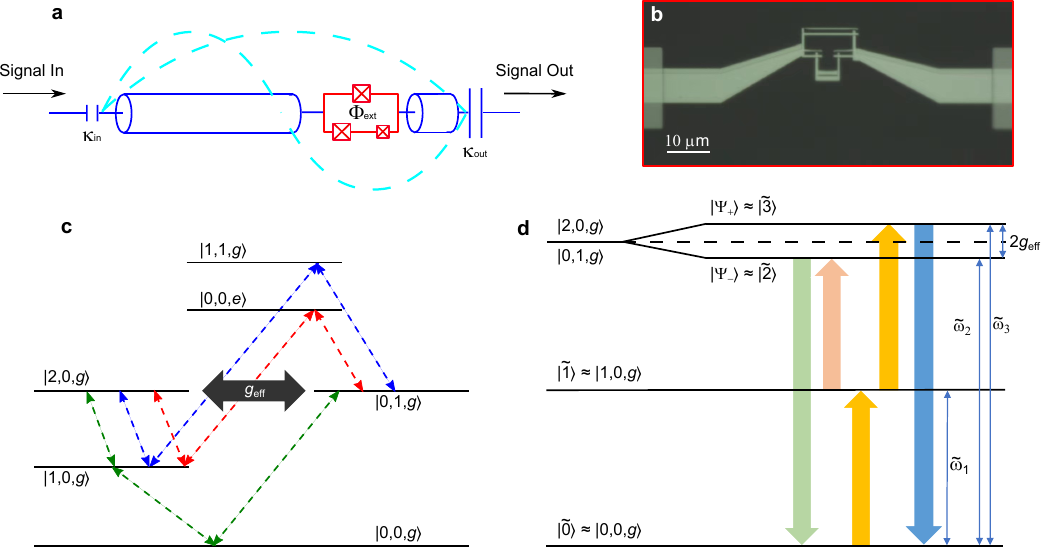}
\caption{\textbf{$|$~Setup.} \textbf{a,}~Schematic of the device. A flux qubit embedded in a ${\lambda}/{2}$ coplanar waveguide resonator, working as a nonlinear coupler between two modes of the resonator. The dashed cyan lines represent the vacuum current distribution of the $n=1$ (${\lambda}/{2}$) and $n=2$ ($\lambda$) modes of the resonator
${\kappa}_{\mathrm{out}}=3{\kappa}_{\mathrm{in}}$, where ${\kappa}_{\mathrm{in(out)}}$ is the loss rate from the input (output) port of the resonator,
\textbf{b,}~Optical image of the flux qubit. \textbf{c,}~Effective coupling mechanism between the bare states $|2,0,g \rangle$ and $|0,1,g \rangle$ via virtual transitions involving intermediate states of the circuit-QED system. Here the first two entries in the kets denote the number of photons in the first two modes of the resonator, and the third indicates the qubit state ($|g\rangle$ is the ground state). {\bf d,}~Scheme of the energy levels at the flux offset corresponding to the minimum gap of the $|2,0,g \rangle-|0,1,g \rangle$ anticrossing, resulting in an effective coupling between these two states. A key feature of this configuration is that, at the minimum anticrossing gap, all the transitions shown by the arrows have comparable large efficiency, with transition matrix elements $|{X}_{1,0}| \simeq |{X}_{+,1}| \simeq |{X}_{+,0}| \sim 1$ (see Supplementary Fig.~2). Moreover, the transition energy $\tilde \omega_3$ is almost equal to $2 \tilde \omega_1$, so that the second harmonic generation (with only two initial photons) and degenerate down conversion (with only one initial photon) can occur {efficiently at sub-photon input levels}.}
\label{fig1}
\end{figure*}
%==============================================

Here, we propose and demonstrate the resonant strong coupling of a two-photon state with a one-photon state in a circuit-QED multimode photonic device. In this system, the ultrastrong light--matter interaction breaks the excitation number conservation and enables higher-order processes via virtual photons, and an external flux bias breaks the parity conservation~\cite{Nori2017PRA, Wang2023NC, Deppe2008NP}. The combined effect of the two enables the strong one--two-photon coupling.

This strong photon-photon interaction induced by the ultrastrong coupling with an off-resonant artificial atom could also be achieved in very different systems and spectral ranges. The key point is to achieve the ultrastrong coupling regime with a single or a few quantum emitters. Strong and even ultrastrong coupling with few organic molecules coupled to a plasmonic nanocavity have been experimentally demonstrated~\cite{Baumberg2016Nature}. Moreover, recently, ab initio calculations have shown that this regime can be achieved even with a single molecule~\cite{Antosiewicz2022ACS}.
Molecules with inversion symmetry would not prevent the coupling of photon states with the same parity (for example $|1\rangle \leftrightarrow |3\rangle$). The significant losses exhibited by plasmonic nanocavities should not prevent the observation of all-optical nonlinearities below the single-photon power level in these nanosystems.

\vspace{.4cm}
\noindent{\bf \large Results}\\
\noindent{\bf Ultrastrongly-coupled circuit-QED device}\\
We realized a superconducting circuit constituted by a coplanar waveguide resonator embedded with an ultrastrongly-coupled flux qubit~\cite{Mooij1999Science, You2007PRB, Chen2017PRA, Wang2020PRApplied, Wang2023NC, Yan2016NC} to implement a multimode system with its two-photon state of the lowest-energy mode strongly coupled to the one-photon state of the second mode (see Methods for details).  Throughout this work, we refer to the $n$th mode as
the $n \lambda/2$-mode. The flux qubit is significantly detuned with respect to the resonant frequencies of the two lowest energy modes of the resonator (dispersive regime), and acts as an effective coupler~\cite{Nori2017PRA, Semba2022PRRe, Brien2021PRL} between the $n =1$ and $n=2$ modes in the absence of any stimulating fields (Fig.~\ref{fig1}c, d). The nonlinear coupling enables two photons in the $n=1$ mode to coherently interact with a single photon in the $n=2$ mode, a hallmark of which is a spectrally resolved mode splitting. We experimentally observed this photon--photon quantum Rabi-like splitting, resulting from the spontaneous hybridization of one- and two-photon states. It indicates that our system operates in the single-photon strong nonlinear coupling regime~\cite{Lukin2014NPh, Bouwmeester2006PRL}, without involving real qubit excitations or any external driving mechanism. As we show here, within this regime, the SHG and stimulated DC processes can occur even when the input signals provide photons to the resonator with mean photon numbers below one.

%==============================================
\begin{figure*}[t]
\includegraphics{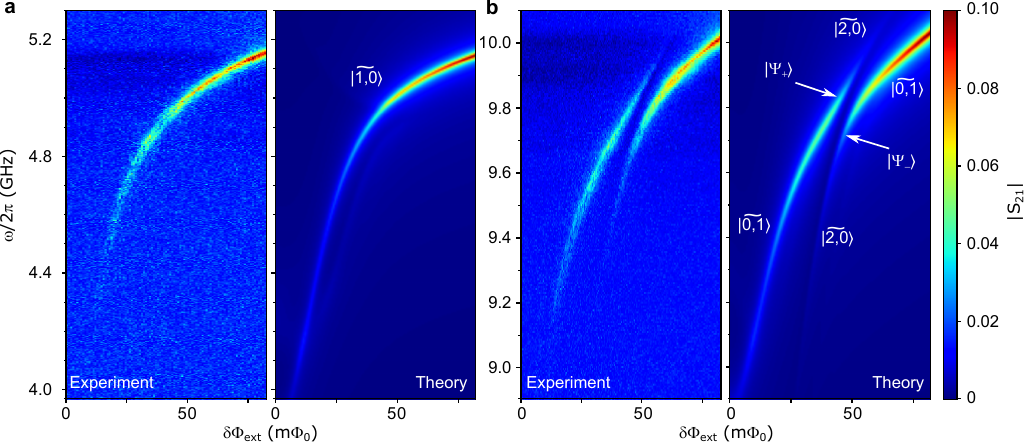}
\caption{\textbf{$|$~Quantum Rabi-like splitting.} Quantum Rabi-like splitting between {a single-photon and a two-photon Fock state}. The left part in each panel (negative flux offset) reports the measured spectra.  The corresponding calculated spectra are reported for positive flux offset (right), using the parity symmetry with respect to the flux offset of the theoretical calculations. \textbf{a,}~Measured and calculated transmission spectra of the one-photon line of the $n=1$ mode of the resonator as a function of the flux offset. \textbf{b,}~Measured and calculated transmission spectra showing the excitation of both the two-photon state of the $n=1$ mode and the one-photon state of the $n=2$ mode of the resonator. An avoided-level crossing between these two lines is clearly visible. From the fitting, we obtain the Rabi frequency for the one--two-photon coupling $g_{\mathrm{eff}}/2\pi = 59 \ \mathrm{MHz}$, the loss rate due to the input--output ports $({\kappa}_{\mathrm{in}} + \kappa_\mathrm{out})/2\pi = 2.6 \ \mathrm{MHz}$, the internal loss rate $\kappa_\mathrm{int} / 2\pi = 10.4 \ \mathrm{MHz}$ (total loss rate of the resonator $\kappa_\mathrm{tot} / 2\pi = (\kappa_\mathrm{in} + \kappa_\mathrm{out} + \kappa_\mathrm{int}) / 2\pi = 13 \ \mathrm{MHz}$), the intrinsic loss rate of the qubit ${\kappa}_{q}/2\pi = 200\ \mathrm{MHz}$, and the pure dephasing rate of the qubit ${\kappa}_{q, \mathrm{dep}}/2\pi = 200\ \mathrm{MHz}$ (the last two have a very weak influence on the calculated spectra).}
\label{fig2}
\end{figure*}
%==============================================

The Hamiltonian of the flux qubit can be written in the basis of two states with persistent currents $\pm I_p$ flowing in opposite directions around the qubit loop as (setting $\hslash =1$) $H_q = -(\Delta \sigma_x+ \varepsilon \sigma_z)/2$, where $\sigma_{x,z}$ are Pauli matrices,
$\Delta$ and $\varepsilon = 2 I_p \delta \Phi_{\rm ext}$ are the tunnel splitting and the  energy bias between the two basis states determined by an adjustable external flux bias $\delta \Phi_{\rm ext}$. The parameters $\Delta /2\pi =12.3\ \mathrm{GHz}$ and $I_p=60\ \mathrm{nA}$ are estimated by numerically diagonalizing the three-junction flux qubit Hamiltonian~\cite{You2007PRB, Yan2016NC}.
The resulting transition frequency between the two energy eigenstates $|g \rangle$ and $|e \rangle$ of the flux qubit is ${\omega}_q=\sqrt{\Delta^2+\varepsilon^2}$.
A schematic of the device and the optical image of the flux qubit are shown in Fig.~\ref{fig1}.

For the flux qubit ultrastrongly coupled to the resonantor, the system can be described by a  generalized two-mode quantum Rabi Hamiltonian~\cite{Nori2017PhyRep}
\begin{equation}
H_{s} = H_q + \sum_{n=1,2} \left[\omega_n a_n^{\dagger}a_n + {g}_n(a_n^{\dagger}+a_n) \sigma_z \right],
\label{Hsys2}
\end{equation}
where ${\omega}_{n}$ are the resonance frequencies of the $n = 1, 2$ modes of the resonator, $a_n^{\dagger}$ and $a_n$ are the corresponding creation and annihilation operators,
and $g_{\mathrm{1,2}}$ are the coupling strengths of the flux qubit with the $n= 1,2$ modes of the resonator. 

In Supplementary Fig.~2, we analyze the influence of higher-energy modes, showing that the two-mode approximation is valid. Owing to the inhomogeneous transmission line geometry due to the qubit presence (see Fig.~\ref{fig1}b), the higher-mode frequency $\omega_2$ is not exactly twice the fundamental resonance
frequency $\omega_1$. Moreover, because of the contribution of the inductance across the qubit loop ($\delta {\mathrm{\Phi}}_{\mathrm{ext}}$-dependent) to the total inductance of the resonator, the resonance frequencies ${\omega}_{1,2}$ of the resonator become V-shaped around the optimal point 
\begin{equation}
{\omega}_{1,2}(\delta {\mathrm{\Phi}}_{\mathrm{ext}})={\omega}_{1,2}(0)/\sqrt{1- \beta_{1,2} \cos (\theta) |\delta {\mathrm{\Phi}}_{\mathrm{ext}}|}, 
\end{equation}
with $\tan (\theta) = \Delta / \varepsilon$ and $\beta_{1,2}$ being constants~\cite{Semba2017NP}. Hereafter, $\tilde \omega_j$ denotes the energy eigenvalues of the Hamiltonian in Eq.~\eqref{Hsys2} with respect to the ground state.

The coupling with the flux qubit determines an effective strong coupling between a single-photon state in mode $2$ ($|0,1,g \rangle$) and the two-photon state of mode $1$ ($|2,0,g \rangle$).
As shown in Fig.~\ref{fig1}c, such a coupling results from higher-order processes involving virtual intermediate transitions enabled by the ultrastrong qubit--resonator interaction~\cite{Nori2019NRP} and, unlike many processes in conventional nonlinear optics, it does not require any external drives.
Such an effective interaction can be written as
\begin{equation}
V_{\mathrm{eff}} = g_{\mathrm{eff}}(|\widetilde{0,1} \rangle \langle \widetilde{2,0}| + |\widetilde{2,0} \rangle \langle \widetilde{0,1}| ),
\label{Heff}
\end{equation}
where $g_{\mathrm{eff}}$ is the nonlinear coupling strength between the two modes (we use ${|\widetilde{n,m} \rangle}$ to indicate photonic states dressed by the interaction with the qubit at higher excitation energy) and is approximately given by an analytic expression~\cite{Nori2017PRA}
\begin{equation}
g_{\rm eff}=\frac{3\sqrt{2}g^2_1g_2{\omega}^2_q\mathrm{sin}(2\theta)\mathrm{cos}(\theta)}{4{\omega}^4_1-5{\omega}^2_1{\omega}^2_q+{\omega}^4_q},
\label{geff}
\end{equation}
when  ${\omega}_2 \simeq {2 \omega}_1$. The effective coupling goes to zero for $\theta = \pi/2$, when the system displays parity symmetry. The effective Hamiltonian in Eq.~(\ref{Heff}) describes the simultaneous annihilation of two photons in the $n=1$ mode and the creation of one photon in the $n=2$ mode as well as the inverse process.
The coupling with the flux qubit introduces further Lamb shifts on the resonance frequencies of the resonator. The qubit population is very small at the one--two-photon avoided level crossing and can be traced out from the dynamics (see Supplementary Fig.~4).

%==============================================
\begin{figure*}[t]
\includegraphics{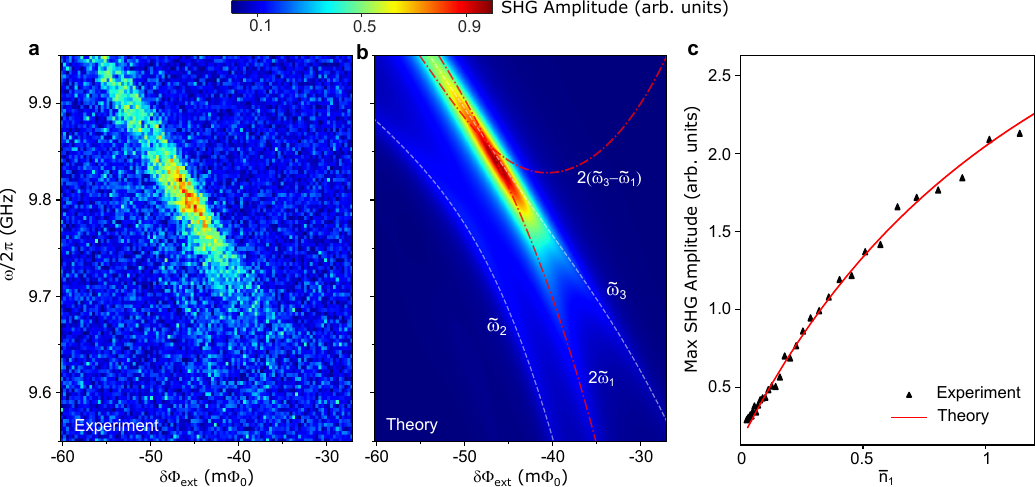}
\caption{\textbf{$|$~Second harmonic generation.} \textbf{a,}~The amplitude of the SHG versus the external flux bias $\delta {\mathrm{\Phi}}_{\mathrm{ext}}$ and the SHG frequency. {The efficiency $\eta \equiv S_{21}^{(2\omega)} / S_{21}^{(\omega)}$ of the SHG is about 0.1 at the point where the SHG amplitude is maximized (${\omega}_1/2\pi=4.9\ \mathrm{GHz}$ and ${\delta \mathrm{\Phi}}_{\mathrm{ext}}=-45~{\rm m \Phi_0}$), which qualitatively agrees with the theoretical result in Supplementary Fig.~2.} \textbf{b,}~Theoretical calculation corresponding to the plot in a. \textbf{c,}~The amplitude of the SHG versus the average photon number ${\overline{n}}_1$ in the resonator. The signal frequency applied at the $n=1$ mode of the resonator is ${\omega}_1/2\pi=4.9\ \mathrm{GHz}$ (at ${\delta \mathrm{\Phi}}_{\mathrm{ext}}=-45~{\rm m \Phi_0}$). The red solid curve is the theoretical fit. {The average photon number in c is determined by contrasting the experimental data---plotting the maximum second-harmonic generation (SHG) amplitude against input power---with the theoretical simulation outlined in Supplementary Eq.~(35), as depicted in c. The power applied in a and b corresponds to an average photon number ${\overline{n}}_1 \simeq 0.25$ in the resonator. }}
\label{fig3}
\end{figure*}
%==============================================

\vspace{.4cm}
\noindent{\bf Transmisison spectra}\\
The device is placed inside a dilution refrigerator to be cooled down to a temperature of 20 mK. Considering the relevant frequencies (${\omega}_1/2\pi\ $around $5.0\ \mathrm{GHz}$, ${\omega}_2/2\pi\ $around $10.0\ \mathrm{GHz}$), the system nearly stays in its ground state at such a low temperature. To see the coupling between the two modes of the resonator, we measured the transmission spectra by applying a weak probe tone with its frequency scanning across the lowest transition energies $\tilde \omega_j$ with $j =1,3$. The spectra are fitted with the numerically calculated excitation energies of the system Hamiltonian $H_{\mathrm{s}}$. The parameters are determined to be 
\begin{subequations}
\begin{align}
{\omega}_1/2\pi = & \ 5.0/\sqrt{1-\beta_1 \cos (\theta) |\delta {\mathrm{\Phi}}_{\mathrm{ext}}|}~\mathrm{GHz},\\ 
{\omega}_2/2\pi = & \ 9.7/\sqrt{1-\beta_2 \cos (\theta) |\delta {\mathrm{\Phi}}_{\mathrm{ext}}|}~\mathrm{GHz},\\
g_1/2\pi = & \ 2.815~\mathrm{GHz},\\
g_2/2\pi = & \ 2.180~\mathrm{GHz},
\end{align}
\end{subequations}
with $\beta_1 = 0.775~\mathrm{(\Phi_0)^{-1}}$ and $\beta_2 = 0.919~\mathrm{(\Phi_0)^{-1}}$. The coupling strengths 
\begin{subequations}
\begin{align}
{g_1}/{\omega_1}= & \ 0.563 > 0.1,\\
{g_2}/{\omega_2}= & \ 0.225 > 0.1
\end{align}
\end{subequations}
at the optimal point (${\delta \mathrm{\Phi}}_{\mathrm{ext}}=0$) provide evidence our system is in the multimode ultrastrong coupling regime. Numerical estimates of coupling strengths from circuit quantization qualitatively agree with the fitted values above (see Supplementary Information). Parameters from the fit are used as input for density matrix calculations which are then compared with the measured spectra. Theoretical spectra throughout this work are obtained by using a generalized master equation \cite{Petruccione2002Theory, Settineri2018Dissipation}, where the interaction with the environment takes into account the hybridization of the subsystems due to the ultrastrong coupling (see Methods for details).

Figure~\ref{fig2} reports the measured (left) and calculated (right) transmission spectra versus the flux offset. The calculated spectra are proportional to the time derivative of the expectation value of the two-mode field momentum operator: 
\begin{equation}
{X} = i ({a}_1 - {a}_1^\dagger) + i \sqrt{\omega_2 / \omega_1} ({a}_2 - {a}_2^\dagger)
\end{equation}
(see Supplemeantary Information for details).
Figure~\ref{fig2}a shows the lowest energy transition $\tilde \omega_1$ corresponding to the excitation of the state 
\begin{equation}
| E_1 \rangle \equiv  |\widetilde{1,0} \rangle\simeq |1,0,g \rangle. 
\end{equation}
Figure~\ref{fig2}b shows the transitions at $\tilde \omega_2$ and $\tilde \omega_3$.
When the two-photon resonance frequency of the $n =1$ mode of the resonator  gets across the single-photon resonance frequency of the $n = 2$ mode ($\omega_{2} \simeq 2 \omega_1$), a clear avoided crossing, induced by the effective interaction in Eq.~\eqref{Heff}, is observed in Fig.~\ref{fig2}b, indicating that our system enters the single-photon strong nonlinear coupling regime~\cite{Bouwmeester2006PRL} $g_{\mathrm{eff}}>\kappa_{\mathrm{tot}}$, where ${\kappa}_{\mathrm{tot}}$ is the total loss rate of the resonator. The nonlinear coupling strength $g_{\mathrm{eff}}/ 2\pi$ extracted from the fit is 59~MHz (half the separation between the double peaks at the minimum splitting), which qualitatively agrees with the perturbative result given by Eq.~\ref{geff} (47~MHz).
While for flux offsets outside the two-modes resonance condition, the system eigenstates are approximately $|2,0,g \rangle$ and $|0,1,g \rangle$; at the anticrossing they are in good approximation their symmetric and antisymmetric superpositions 
\begin{equation}
|\psi_\pm \rangle =|2,0,g \rangle \pm |0,1,g \rangle. 
\end{equation}
Note that outside the avoided-level crossing in Fig.~\ref{fig2}b, the spectral line corresponding to the excitation of the two-photon state $|2,0,g \rangle$ is poorly excited by the weak probe tone.
Two-photon contributions get significantly excited only owing to their hybridization with one-photon states at the avoided crossing. The emission of photon pairs under very weak excitation is maximized in this region (see Supplementary Fig.~3).
Theoretical calculations accurately reproduce all the spectra in Fig.~\ref{fig2}b, thus confirming our interpretation of the data.
A peculiar feature of this extreme quantum nonlinear regime is that linear spectra in Fig.~\ref{fig2}b  are able to evidence quantum nonlinear optical processes.

%==============================================
\begin{figure*}[t]
\includegraphics[width = \linewidth]{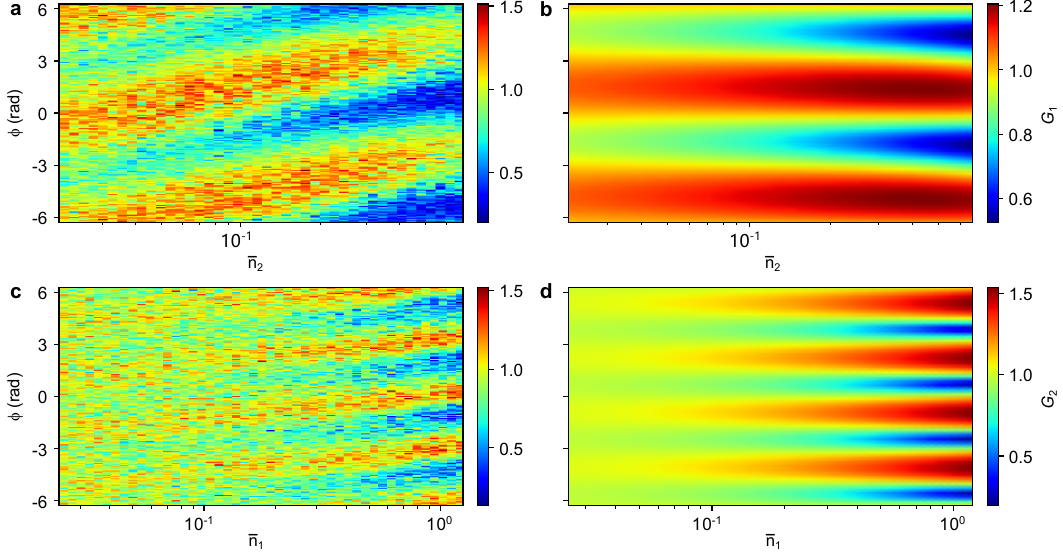}
\caption{\textbf{$|$~Interference between two probe tones.} \textbf{a(c),}~Gain of the transmitted amplitude of the signal field in the $n=1$ ($n=2$) mode of the resonator versus the average photon number ${\overline{n}}_{\mathrm{2(1)}}$ and the phase of the control field in the $n=2$ ($n=1$) mode of the resonator. The average photon numbers of the signal fields in the resonator's $n=1$ and $n=2$ modes are about 0.25 and 0.13, respectively. The frequencies of the probe tones are $\omega/ 2\pi =4.905$ GHz and $2 \omega/ 2\pi$ (at ${\delta \mathrm{\Phi}}_{\mathrm{ext}}=-46~{\rm m \Phi_0}$). \textbf{b (d),}~Theoretical calculations corresponding to the plots in a (c). {The gain is defined as the ratio between the output signal amplitude at $\omega$ ($2 \omega$) when a secondary signal is applied at $2 \omega$ ($\omega$), and the output amplitude of the same signal at $\omega$ ($2 \omega$) in the absence of the secondary signal applied at $2 \omega$ ($\omega$).} The phase shifts in the experimental data in a (c) are an artifact due to an electronic delay.}
\label{fig4}
\end{figure*}
%==============================================

\vspace{.4cm}
\noindent{\bf Nonlinear optics below the single-photon power level}\\
To further characterize the nature of the strong nonlinear coupling, we measured the SHG at the single-photon power level by applying a weak signal at $\omega$ and monitoring the SHG signal at $2 \omega$. The result is shown in Fig.~\ref{fig3}. As expected, the SHG signal shown in Fig.~\ref{fig3}a is more intense near the minimum of the avoided-level crossing and for flux offsets where the system is double resonant: $\tilde \omega_3 = 2 \tilde \omega_1$, as confirmed (despite some shift due to some small discrepancy in the fitted energy levels) by the theoretical calculations in Fig.~\ref{fig3}b (see Supplementary Information for details). The efficiency 
\begin{equation}
\eta \equiv S_{21}^{(2\omega)} / S_{21}^{(\omega)}
\end{equation}
of the SHG signal, shown in Fig.~\ref{fig3}a, is about 0.1 at the point where the SHG amplitude is maximum. Such efficiency is achieved with an input corresponding to only 0.25 photons  into the resonator, clearly operating in the quantum limit. To the best of our knowledge, SHG in the true (without additional many-photon drives) quantum limit has never been experimentally observed until now~\cite{Yanagimoto2022Optica}.
A theoretical estimate (see Supplementary Fig.~3) shows that the efficiency can approach $\eta = 0.3$ for 1.5 input photons. The efficiency can be significantly improved by reducing the internal losses of the resonator and setting the system at the perfect feeding condition so that no output signal at $\omega$ is present~\cite{Bonizzoni2025}.
Also, to the best of our knowledge, near-deterministic photon up/down conversion has so far been achieved predominantly using atomic level transitions~\cite{Inomata2014PRL}. Crucially, our approach has the potentiality to achieve near-deterministic pure photon up/down conversion without involving atomic transitions, thereby avoiding the associated losses.

Figure~\ref{fig3}c shows that the SHG amplitude is thresholdless and starts with a linear dependence on the mean value of input photons $\bar n_1$, while it tends to saturate for higher input powers. It can clearly be distinguished from the noise floor even for the mean number of input photons in the resonator ${\overline{n}}_1$ significantly below $1$, which is a striking unprecedented result, made possible via the photonic strong coupling regime shown in Fig.~\ref{fig2}b. The mean number of input photons in the resonator is evaluated using a standard input--output relationship~\cite{Milburn2008Springer}(see Supplementary Information for details). {During the calibration of the photon number, the obtained fitted parameters allowed us to reproduce in very good agreement not only the data in Fig.~\ref{fig3}c but also the spectra in Fig.~\ref{fig2} and the interference fringes in Fig.~\ref{fig4}. Hence, although there might be a slight deviation from the true photon number, we expect it to be less than $10~\%$ of our fitted values.} The hybridization between one- and two-photon states at the minimum of the avoided-level crossing gives rise to matrix elements ${X}_{+,1},\, {X}_{+,0},\, {X}_{1,0}$ all of the order of unity, which enable observation of the SHG well below the single-photon power level.

A further direct consequence of the achieved photonic strong coupling  is the observation of interference between a probe at $\omega$ ($2 \omega$) and a control tone at $2 \omega$ ($\omega$) both at the single-photon power level, resulting in the amplification or suppression of the transmitted amplitude of probe signals~\cite{Astafiev2010PRL}. The amplification and suppression are phase dependent, just as in the case of degenerate parametric amplifiers~\cite{Tsai2008APL, Lehnert2008NP}. The difference is that a much stronger pump tone is not needed here, and we observe these interference phenomena at an incredibly low-power limit (average photon number in the resonator below one).
The results are shown in Fig.~\ref{fig4}. Changing the phase of the control tones, we can see that the transmitted amplitude of the signal at the $n=2$ ($n=1$) mode (signal field) is amplified or suppressed periodically with visibility approaching 1.
It turns out that the visibility of the interference  for the signal at $2 \omega$ becomes relevant for higher intensities of the control signal (at $\omega$) (see Fig.~\ref{fig4}c,d) with respect to what is observed  for the signal at $\omega$ (see Fig.~\ref{fig4}a,b).
Conversely, to obtain a high visibility for interference at $\omega$, it requires a larger probe signal at $\omega$. This difference originates from the fact that the interference at $2 \omega$ requires the stimulated conversion of two photons of the control tone. On the contrary, the probe signal at $\omega$ receives {two photons} from the down conversion of photons at $2 \omega$ of the control tone. The observed tilting in the experimental data is attributed to a slight change in the phase of the signal when adjusting the output power. These two-tone interference data are well reproduced (except, of course, for the tilting) by numerical calculations based on the generalized master equations including a two-tone interaction term [see Supplementary Eq.~(36)], shown in Fig.~\ref{fig4}b,d. 

We note that this work does not provide any experimental evidence of spontaneous down-conversion. This is a direct consequence of our setup, which selects only coherent signals, with the phase determined by the input signal. This technique allows to filter thermal noise and to allow the detection of very small signals. Spontaneous down conversion is not a coherent signal and it is thus filtered out. However, Fig.~\ref{fig4}a provides clear evidence of  stimulated down-conversion (phase-dependent amplification of the signal at $\omega$ in the presence of a tone at $2 \omega$), which gives rise to a coherent signal that can be detected by our setup. A simple understanding of the phase sensitivity of the signal amplification at $\omega$ can be easily obtained considering the standard evolution equation for the linear parametric amplifier in the weak pump limit (see Supplementary Eq.~40).

\vspace{.4cm}
\noindent{\bf \large Discussion}\\
We experimentally observed an unstimulated strong coupling between {a single-photon and a two-photon Fock state}. The observed single-photon strong nonlinear coupling is enabled by an embedded-in flux qubit, which, while remaining in its ground state, acts as an effective coupler between two modes of a superconducting coplanar waveguide resonator. The strong coupling between photon states of opposite parity is enabled by the spontaneous mixing of even- and odd-number photon states, which is a unique feature of the generalized quantum Rabi Hamiltonian in the ultrastrong coupling regime~\cite{Nori2017PRA, Wang2023NC}.

The counterpart in time-resolved experiments of the quantum Rabi splitting reported here are quantum Rabi oscillations enabling the conversion of a single photon into {two photons} and vice versa with a period $T_R = 2 \pi / g_{\rm eff}$ and with an efficiency that can potentially approach $100\%$~\cite{Nori2017PRA} (for $g_{\rm eff}$ significantly larger than the losses). In contrast to most conventional implementations of nonlinear optics, the regime demonstrated here can reach unit efficiency only using a minimal number of photons and without exciting atomic transitions, which introduce additional losses. For example, in many cases pure dephasing in superconducting artificial atoms is the main source of decoherence.

These results unlock a new regime for photons in quantum nonlinear optics~\cite{Lukin2014NPh, Bouwmeester2006PRL} (coherent interaction between photon multiplets), which has important implications in both fundamental research~\cite{Nori2017PRA} and applications~\cite{Zeilinger2011Nature, Niu2018PRL}. Examples of direct applications include 
\begin{enumerate}[(i)]
\item the realization of deterministic quantum light (single photon and photon pair) sources as in the cases of traditional spontaneous parametric down conversion (SPDC) but with an unprecedented efficiency (approaching $100\%$); 
\item an all-optical deterministic controlled-phase gate ($|\widetilde{0,1}\rangle \leftrightarrow -|\widetilde{0,1}\rangle$, a geometric phase effect under a full Rabi oscillation ($2 T_R$), where the presence or the absence of a photon in mode 1 corresponds to the control qubit). 
\end{enumerate}
This two-photon quantum phase gate represents a key element for the realization of a still elusive universal nonlinear optical quantum computing (NOQC)~\cite{Zeilinger2011Nature, Niu2018PRL}. These controlled quantum Rabi oscillations can be realized by using suitable flux-offset pulses to tune the system in and out the minimum of the avoided level crossing $|\widetilde{0,1}\rangle \leftrightarrow |\widetilde{2,0}\rangle$.

\vspace{.4cm}
\noindent{\bf \large Methods}\\
\noindent{\bf Superconducting circuit device}\\
Our device is a 3 mm $\times $ 10 mm chip fabricated on a single crystal sapphire substrate (see Supplementary Fig.~1b). The superconducting coplanar waveguide resonator is defined by patterning a niobium thin film using optical lithography. The flux qubit consists of two identical large JJs and a smaller JJ (with an area ratio $\alpha \approx \mathrm{0.5}$) in a loop geometry. 

The qubit loop and its connection to the center conductor of the coplanar waveguide resonator are made by electron beam lithography and double-angle evaporation of aluminum. Placing the qubit loop near the antinode of the current distribution for the $n=2$ mode of the resonator makes the qubit coupled to the $n=1$ mode and the $n=2$ mode of the resonator simultaneously (see Fig.~\ref{fig1}a). 

The values of the Josephson energy $E_J/h=173\ \mathrm{GHz}$, the area ratio $\alpha \approx \mathrm{0.5}$ and ${E_J}/{E_C} \approx 100$ of the JJs in the qubit are optimized to ensure the flux qubit working as an effective nonlinear coupler, i.e., both far-away detuned and strongly coupled with the relevant modes of the resonator, and are used to numerically diagonalize the three-junction flux qubit Hamiltonian to estimate the values of $\Delta$ and $I_p$.

The device is coupled to the environment through two capacitors of different sizes (see Fig.~\ref{fig1}). 
One of these input-output ports is used as an input port to excite the device with one or two coherent weak tones, while the other port is used to detect signals. In order to reduce the output through the input port, the size of the output capacitor has been chosen to be larger ($\simeq 1.7$ times) than that of the input one. Notice that the loss rate associated to a capacitive port is proportional to the square of the corresponding capacitance.

\vspace{.4cm}
\noindent{\bf Input--output and generalized master equation}\\
The standard quantum optics approach to describe open quantum systems no longer holds in the case of ultrastrong coupling. To properly derive the time evolution of the system and the input--output relations, the interaction term between the system and the environment has to be expressed on the basis of the eigenstates of the system. Moreover, the field input--output relations include anharmonic operators. This problem can be circumvented by expressing the time evolution of the field operators in terms of the generalized master equation, rather than the nonlinear input--output relations (See Supplemeantary Information for details).

\vspace{.4cm}
\noindent{\bf \large Data availability}\\
The data that support the findings of this study are available at Code Ocean (https://doi.org/10.24433/CO.4816683.v1~\cite{CO}).

\vspace{.4cm}
\noindent{\bf \large Code availability}\\
The code that support the findings of this study are available at Code Ocean (https://doi.org/10.24433/CO.4816683.v1~\cite{CO}).

\vspace{.4cm}
\noindent{\bf \large Acknowledgments}

\begin{acknowledgments}
This work is supported by the National Key Research and Development Program of China (Grant No.~2022YFA1405200), the National Natural Science Foundation of China (Grants No.~92265202, No.~92365210, No.~62074091, No.~12574547, No.~U2230402, No.~12374325, and No.~12304387), and Tsinghua University Initiative Scientific Research Program.
A.R. acknowledges the QuantERA grant SiUCs (Grant No.~731473), the PNRR MUR project PE0000023-NQSTI, and the ICSC - Centro Nazionale di Ricerca in High-Performance Computing, Big Data and Quantum Computing.
Y.L. acknowledges the support of Beijing Municipal Science and Technology Commission (Grant No.~Z221100002722011), and Young Elite Scientists Sponsorship Program by CAST (Grant No.~2023QNRC001).
S.S. acknowledges the Army Research Office (ARO) (Grant No.~W911NF1910065).
F.N. is supported in part by: the Japan Science and Technology Agency (JST) [via the CREST Quantum Frontiers program Grant No. JPMJCR24I2, the Quantum Leap Flagship Program (Q-LEAP), and the Moonshot R\&D Grant No. JPMJMS2061].
\end{acknowledgments}

\vspace{.4cm}
\noindent{\bf \large Author contributions}\\
S.-P.W., T.L. and J.Q.Y. conceived the experiment. Y.W., M.C. and W.W. fabricated the sample. S.-P.W. performed the experiment. S.-P.W., A.M., A.R., and S.S. analyzed and interpreted the data. A.M., A.R., F.N., and S.S. developed the theory. S.-P.W., A.M., A.R., Y.L., H.S., T.L., F.N., S.S. and J.Q.Y. discussed the results and contributed to the writing of the manuscript.

\vspace{.4cm}
\noindent{\bf \large Competing Interests}\\
The authors declare no competing interest.

\vspace{.4cm}
\noindent{\bf \large Additional information}\\
Correspondence and requests for materials should be addressed to Tiefu Li, Salvatore Savasta, or J. Q. You.

\end{document}

\renewcommand\figurename{Supplementary Fig.}

\title{Supplementary Information for \textquotedblleft{Strong coupling between {a single-photon and a two-photon Fock state}}\textquotedblright{}}

\author{Shuai-Peng Wang}
\affiliation{Beijing Academy of Quantum Information Sciences, Beijing 100193, China}
\affiliation{Quantum Physics and Quantum Information Division, Beijing Computational Science Research Center, Beijing 100193, China}
\affiliation{Zhejiang Key Laboratory of Micro-Nano Quantum Chips and Quantum Control, School of Physics, and State Key Laboratory for Extreme Photonics and Instrumentation, Zhejiang University, Hangzhou 310027, China}

\author{Alberto Mercurio}
\affiliation{Institute of Physics, {\'E}cole Polytechnique F{\'e}d{\'e}rale de Lausanne (EPFL), CH-1015 Lausanne, Switzerland}
\affiliation{Center for Quantum Science and Engineering, {\'E}cole Polytechnique F{\'e}d{\'e}rale de Lausanne (EPFL), CH-1015 Lausanne, Switzerland}

\author{Alessandro Ridolfo}
\affiliation{Dipartimento di Fisica e Astronomia, Universit{\`a} di Catania, 95123 Catania, Italy}

\author{Yuqing Wang}
\affiliation{Beijing Academy of Quantum Information Sciences, Beijing 100193, China}

\author{Mo Chen}
\affiliation{Beijing Academy of Quantum Information Sciences, Beijing 100193, China}

\author{Wenyan Wang}
\affiliation{Beijing Academy of Quantum Information Sciences, Beijing 100193, China}

\author{Yulong Liu}
\affiliation{Beijing Academy of Quantum Information Sciences, Beijing 100193, China}

\author{Huanying Sun}
\affiliation{Beijing Academy of Quantum Information Sciences, Beijing 100193, China}

\author{Tiefu Li}
\email{litf@tsinghua.edu.cn}
\affiliation{School of Integrated Circuits, and Frontier Science Center for Quantum Information, Tsinghua University, Beijing 100084, China}
\affiliation{Beijing Academy of Quantum Information Sciences, Beijing 100193, China}

\author{Franco Nori}
\affiliation{Quantum Computing Center, RIKEN, Wako-shi, Saitama 351-0198, Japan}
\affiliation{Physics Department, The University of Michigan, Ann Arbor, Michigan 48109-1040, USA}

\author{Salvatore Savasta}
\email{salvatore.savasta@unime.it}
\affiliation{Dipartimento di Scienze Matematiche e Informatiche, Scienze Fisiche e Scienze della Terra, Universit{\`a} di Messina, I-98166 Messina, Italy}

\author{J. Q. You}
\email{jqyou@zju.edu.cn}
\affiliation{Zhejiang Key Laboratory of Micro-Nano Quantum Chips and Quantum Control, School of Physics, and State Key Laboratory for Extreme Photonics and Instrumentation, Zhejiang University, Hangzhou 310027, China}

%\date{\today}

%%%%%%%%%%%%%%%%%%%%%%%%%%%%%%%%%%%%%%%%%%%%%%%%%%%%%%%%%

\maketitle

\section{Experimental Setup}

\noindent
The experimental setup is shown in Supplementary Fig.~\ref{fig:S1}. The transmission spectra of the multimode ultrastrongly-coupled qubit--resonator system in Fig.~2 of the main text are measured with a vector network analyzer. A signal generator and a spectrum analyzer are used for the SHG and the interference measurements in Fig.~3 and Fig.~4 of the main text. 

A current source is used to generate the external flux bias $\delta \Phi_{\rm ext}$ via an on-chip control line. The input signals are all attenuated and filtered at different temperature stages in the dilution refrigerator before reaching the sample. The output signals are amplified by both a cryogenic and a room temperature amplifier before being collected by the network analyzer and the spectrum analyzer. Isolators and low-pass filters (LPF) are used to protect the sample from the amplifiers' noise. 

The data presented in Figs.~2, 3, and 4 in the main text were all acquired as coherent signals using a vector network analyzer (VNA). For Fig.~2 in the main text, the VNA was employed to measure the S21 parameter of the transmitted signal. In the case of Fig.~3 in the main text, the VNA detected the SHG signal emanating from the device, which was excited by an external signal generator. The VNA’s output port was terminated, and it functioned solely to capture the SHG signal. For Fig.~4 in the main text, the VNA measured the S21 parameter of the signal at the frequency $\omega/2\omega$, while a simultaneous signal at $2\omega/\omega$ frequency was applied from a separate signal generator.

%========================================================================
\begin{figure*}[hbt!]
	 	\centering
	 	\includegraphics[width=\linewidth]{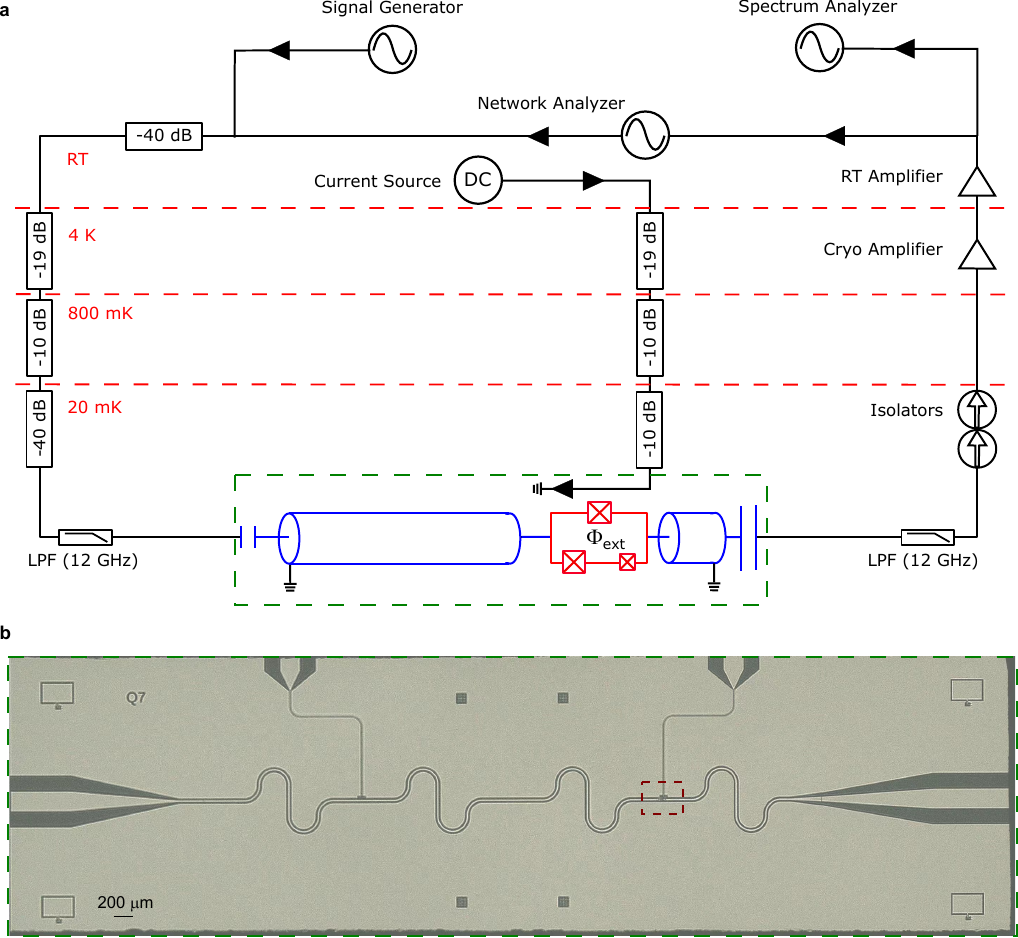}
		\caption{{\bf a,}~Schematic of the experimental setup. RT: room temperature. Cryo: cryogenic. LPF: low pass filter. The green dashed rectangle indicates the region of the chip. {\bf b,}~The optical image of the chip. The red dashed rectangle indicates the location of the flux qubit.}
		\label{fig:S1}
\end{figure*}
%========================================================================

\section{Theory}
\noindent {\bf Circuit quantization}\\
To derive the total Hamiltonian of the system, here we follow the procedure applied in Ref.~\cite{Bourassa2012Josephson}, where the quantization was applied to the case of a single Josephson junction or a superconducting quantum interference device (SQUID) embedded in a transmission line, while here we consider the flux qubit. A similar procedure can also be found in Ref.~\cite{Peropadre2013Nonequilibrium}. The flux qubit circuit is composed by a loop of three Josephson junctions, as in Supplementary Fig.~\ref{fig:S1}. Without the qubit, the Lagrangian of the bare transmission line is given by~\cite{Blais2021Circuit}
\begin{equation}
    \mathcal{L}_{\mathrm{tl}} = \int_{-l}^{l} \left[ \frac{c_0}{2} \dot{\psi}^2 (x, t) - \frac{\left[ \partial_x \psi (x, t) \right]^2}{2 l_0} \right] dx \, ,
\end{equation}
where $c_0$ and $l_0$ are the capacitance and inductance per unit length, respectively, and $\psi (x, t)$ is the flux along the waveguide. For simplicity, we consider $c_0$ and $l_0$ as constants, although the inhomogeneous geometry of the waveguide in the region of the qubit can make them space-dependent~\cite{Bourassa2009Ultrastrong}. The qubit is placed at the point $x_0$, and the related Lagrangian is
\begin{equation}
    \begin{split}
        \mathcal{L}_\mathrm{q} = & \frac{C_j}{2} \dot{\varphi}^2 + \frac{C_j}{2} \dot{\varphi_1}^2 + \frac{ \alpha C_j}{2} \dot{\varphi_2}^2 \\
        &+ E_j \cos \left( \frac{\varphi}{\varphi_0} \right) + E_j \cos \left( \frac{\varphi_1}{\varphi_0} \right) \\
        &+  \alpha E_j \cos \left( \frac{\varphi_2}{\varphi_0} \right) \, ,
    \end{split}
\end{equation}
where $C_j$ and $E_j$ are the capacitance and energy of the junction, $\varphi_0 = \hbar / (2 e)$ is the reduced flux quantum, $\varphi_1$ and $\varphi_2$ are the fluxes across the junctions in the lower arm, while $\varphi = \psi (x_0^+, t) - \psi (x_0^-, t)$ is the flux across the junction in the upper arm.

By moving the quadratic term of $\cos (\varphi / \varphi_0)$ into the resonator term, the total Lagrangian can be expressed as
\begin{equation}
    \mathcal{L} = \mathcal{L}_\mathrm{tl}^\prime + \mathcal{L}_\mathrm{q}^\prime \, ,
\end{equation}
where the prime in the resonator term indicates that it includes the quadratic term $E_j \varphi^2 / (2 \varphi_0^2)$, while the prime in the qubit term indicates that it does not include this term.

To proceed with the quantization, we need to express the resonator Lagrangian in terms of the normal modes. This can be done by solving the Euler-Lagrange equation
\begin{equation}
    \sum_{\nu = x,t} \partial_\nu \left( \frac{\partial \mathcal{L}_\mathrm{tl}^\prime}{\partial \left[ \partial_\nu \psi (x, t) \right]} \right) - \frac{\partial \mathcal{L}_\mathrm{tl}^\prime}{\partial \psi (x, t)} = 0 \, ,
\end{equation}
together with the boundary conditions at the two ends of the resonator and the constraint dictated by the qubit at position $x_0$. We now express the field in terms of the modes
\begin{equation}
    \psi (x, t) = \sum_n \psi_n (t) u_n (x) \, ,
\end{equation}
with $\ddot{\psi}_n (t) = -\omega_n^2 \psi_n (t)$ a function of time oscillating at mode frequency $\omega_n$ and $u_n (x)$ are determined by the boundary conditions. We define these functions to be of the form
\begin{equation}
    \label{eq-u_n-modes}
    u_n (x) = A_n \begin{cases}
        \sin \left[ k_n \left( x + l \right) - \phi_n \right] \ \text{ for } -l \leq x < x_0^- \, , \\
        B_n \sin \left[ k_n \left( x - l \right) + \phi_n \right] \ \text{ for } x_0^+ < x \leq l \, ,
    \end{cases}
\end{equation}
with $k_n = \omega_n \sqrt{l_0 c_0}$, $A_n$, $B_n$ and $\phi_n$ are constants determined by the various constraints. The boundary conditions at the two ends of the resonator are
\begin{equation}
    \frac{1}{l_0} \frac{\partial \psi (x, t)}{\partial x} \bigg|_{x = -l, l} = 0 \, ,
\end{equation}
which imply $\phi_n = \pi / 2$. The current on both sides of the qubit is equal, giving the condition
\begin{equation}
    \label{eq-constraint-qubit}
    \frac{1}{l_0}  \frac{\partial \psi (x, t)}{\partial x} \bigg|_{x = x_0^-} = \frac{1}{l_0}  \frac{\partial \psi (x, t)}{\partial x} \bigg|_{x = x_0^+} = C_j \ddot{\varphi} + \frac{\varphi}{L_j} \, ,
\end{equation}
with $L_j = \varphi_0^2 / E_j$ the flux-dependent Josephson inductance. The first equation gives the value of $B_n$
\begin{equation}
    B_n = \frac{\cos \left[ k_n \left( x_0 + l \right) - \phi_n \right]}{\cos \left[ k_n \left( x_0 - l \right) + \phi_n \right]} \, .
\end{equation}
The value of $k_n$ is found by inserting the mode shape in Supplementary Eq.~\eqref{eq-u_n-modes} into the constraint in Supplementary Eq.~\eqref{eq-constraint-qubit}, giving the following transcendental equation
\begin{equation}
    \begin{split}
        & \left\{ \tan \left[ k_n \left( x_0 - l \right) + \phi_n \right] - \tan \left[ k_n \left( x_0 + l \right) - \phi_n \right] \right\} \\
        & \times \left[ \frac{l_0 l}{L_j} - \left( k_n l \right)^2 \frac{C_j}{c_0 l} \right] - k_n l = 0 \, .
    \end{split}
\end{equation}
The profile shapes $u_n (x)$ satisfy the orthonormality condition~\cite{Bourassa2012Josephson}
\begin{equation}
    \langle u_n(x) \cdot u_m (x) \rangle = C_\Sigma \delta_{nm} \, ,
\end{equation}
with $C_\Sigma = 2 l c_0 + C_j$.

Using the orthonormality condition above and the mode decomposition in Supplementary Eq.~\eqref{eq-u_n-modes}, we can express the total Lagrangian in terms of the normal modes
\begin{equation}
    \mathcal{L} = \sum_n \left( \frac{C_\Sigma}{2} \dot{\psi}_n^2 - \frac{\psi_n^2}{2 L_n} \right) + \mathcal{L}_\mathrm{q} \, ,
\end{equation}
with $L_n = \omega_n^2 C_\Sigma$.

When applying an external field $\Phi_\mathrm{ext}$, the three junction fluxes are related by the flux quantization
\begin{equation}
    \varphi + \varphi_1 + \varphi_2 = \Phi_\mathrm{ext} \, .
\end{equation}
Thus, we can express $\varphi_2$ in terms of $\varphi$, $\varphi_1$ and $\Phi_\mathrm{ext}$. We now pass to the Hamiltonian formalism, obtaining
\begin{equation}
    \begin{split}
        H = & \sum_n \left( \frac{q_n^2}{2 C_\Sigma} + \frac{\psi_n^2}{2 L_n} \right) + \frac{\varphi^2}{2 L_j} + \frac{q_1^2}{2 C_j} \\
        &- E_j \cos \left( \frac{\varphi}{\varphi_0} \right) - E_j \cos \left( \frac{\varphi_1}{\varphi_0} \right) \\
        &- \alpha E_j \cos \left( \frac{\Phi_\mathrm{ext} - \varphi - \varphi_1}{\varphi_0} \right) \, ,
    \end{split}
\end{equation}
where $\varphi = \sum_n \psi_n \Delta u_n$ is a resonator variable, with $\Delta u_n = u_n (x_0^+) - u_n (x_0^-)$.

By using the trigonometric property $\cos(x + y) = \cos x \cos y - \sin x \sin y$ and the last term, with $x = (\varphi - \Phi_\mathrm{ext}) / \varphi_0$ and $y = \varphi_1 / \varphi_0$, we can express the Hamiltonian in the form
\begin{equation}
    \label{eq-hamiltonian-classical}
    \begin{split}
        H = & \sum_n \left( \frac{q_n^2}{2 C_\Sigma} + \frac{\psi_n^2}{2 L_n} \right) + \frac{\varphi^2}{2 L_j} + \frac{q_1^2}{2 C_j} \\
        &- \tilde{E}_j (\Phi_\mathrm{ext}) \cos \left( \frac{\varphi}{\varphi_0} - \delta_0 \right) - E_j \cos \left( \frac{\varphi_1}{\varphi_0} \right) \\
        &- \alpha E_j \cos \left( \frac{\varphi - \Phi_\mathrm{ext}}{\varphi_0} \right) \left[ \cos \left( \frac{\varphi_1}{\varphi_0} \right) - 1 \right] \\
        &+ \alpha E_j \sin \left( \frac{\varphi - \Phi_\mathrm{ext}}{\varphi_0} \right) \sin \left( \frac{\varphi_1}{\varphi_0} \right) \, ,
    \end{split}
\end{equation}
where
\begin{equation}
    \label{eq-flux-dependent-Ej}
    \tilde{E}_j (\Phi_\mathrm{ext}) = E_j \sqrt{ 1 + \alpha^2 \tan^2 \left( \frac{\Phi_\mathrm{ext}}{2\varphi_0} \right) } \cos \left( \frac{\Phi_\mathrm{ext}}{2\varphi_0} \right)
\end{equation}
is the flux-dependent Josephson energy, and $\delta_0 = \alpha \tan (\Phi_\mathrm{ext} / 2\varphi_0)$ is the phase shift. By expanding the term $\tilde{E}_j (\Phi_\mathrm{ext}) \cos \left( \frac{\varphi}{\varphi_0} - \delta_0 \right)$ up to the second order, we obtain a flux-dependent renormalization of the bare modes frequencies $\omega_n^\prime (\Phi_\mathrm{ext})$, together with a flux-dependent self- and cross-Kerr~\cite{Bourassa2012Josephson}. Here we neglect the effect of the Kerr terms, as they don't play a significant role in the dynamics studied in this work, while the renormalization of the frequency of the modes gives the V-shape behavior. A similar flux-dependent renormalization of the modes frequencies is also present in a flux qubit interacting with a $LC$ resonator~\cite{Yoshihara2017Superconducting}.

%========================================================================
\begin{figure}[hbt!]
    \centering
    \includegraphics[width=\linewidth]{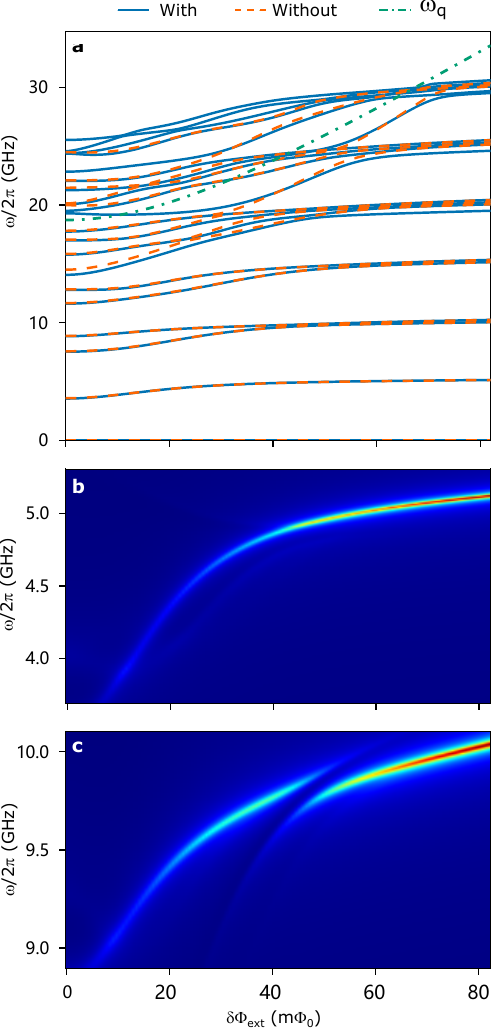}
    \caption{Influence of the fourth mode of the resonator. {\bf a,} eigenvalues of the total Hamiltonian (solid blue) including the fourth mode with coupling $g_4 / 2\pi = 2 \ \mathrm{GHz}$, without the fourth mode (dashed orange) and bare qubit frequency (dash-dotted green) as a function of the flux bias $\delta\Phi_\mathrm{ext}$. The qubit becomes resonant with the fourth mode at $\delta\Phi_\mathrm{ext} \simeq 40 ~ {\rm m}\Phi_0$. However, the low energy eigenvalues remain unperturbed. {\bf b, c} transmission spectra for the first and second mode, respectively. The results are indentical to the ones presented in the main text.}
    \label{fig:fourth-mode-study}
\end{figure}
%========================================================================

We now quantize the Hamiltonian in Supplementary Eq.~\eqref{eq-hamiltonian-classical}. Furthermore, we can apply the trigonometric property to separate the terms in $\varphi - \Phi_\mathrm{ext}$ in the last two terms, and then expanding them up to first order in $\varphi$, obtaining the following Hamiltonian
\begin{equation}
    \begin{split}
        {H} \simeq & \sum_n \hbar \omega_n^\prime (\Phi_\mathrm{ext}) {a}_n^\dagger {a}_n + \frac{{q}_1^2}{2 C_j} - E_j \cos \left( \frac{{\varphi}_1}{\varphi_0} \right) \\
        & - \alpha E_j \cos \left( \frac{\Phi_\mathrm{ext}}{\varphi_0} \right) \left[ \cos \left( \frac{{\varphi}_1}{\varphi_0} \right) - 1 \right] \\
        & - \alpha E_j \sin \left( \frac{\Phi_\mathrm{ext}}{\varphi_0} \right) \sin \left( \frac{{\varphi}_1}{\varphi_0} \right) \\
        & + \frac{\alpha E_j}{\varphi_0^2} \cos \left( \frac{\Phi_\mathrm{ext}}{\varphi_0} \right) {\varphi} {\varphi}_1 \, .
    \end{split}
\end{equation}
Notice that we have neglected the term $\varphi [\cos ({\varphi}_1/ \varphi_0 ) - 1]$ and considered ${\varphi} \sin ( {\varphi}_1 / \varphi_0 ) \simeq {\varphi} {\varphi}_1 / \varphi_0$. Not doing these approximations would only lead to a small renormalization of the coupling strengths and modes frequencies, when projecting in the two-level subspace.

By projecting the total Hamiltonian into the two lowest-energy levels of the qubit, we get
\begin{equation}
\label{eq-app:Rabi Hamiltonian derivation}
\begin{split}
    {H} =& \frac{\hbar \omega_q}{2} {\sigma}_z + \sum_n \left[\hbar \omega_n^\prime (\Phi_\mathrm{ext}) {a}_n^\dagger {a}_n \right . \\
    &+ \left. \hbar g_n \left( -\sin \theta\, {\sigma}_x + \cos \theta\, {\sigma}_z \right) \left({a}_n + {a}_n^\dagger \right) \right] \, ,
\end{split}
\end{equation}
with
\begin{equation}
    \label{g_n}
    g_n = \cos \left( \frac{\Phi_\mathrm{ext}}{\varphi_0} \right) \frac{\alpha E_j}{\hbar \varphi_0} \sqrt{\frac{\hbar}{2 C_\Sigma \omega_n}} \Delta u_n \, ,
\end{equation}
which is equivalent to the Hamiltonian in Eq.~(1) of the main text, but expressed in the qubit energy basis instead of the persistent current one.  The qubit position corresponds to the current node of the third mode of the waveguide ($\Delta u_3 \approx 0$), leading to $g_3 \approx 0$. Hence, we can consider only the first and second modes for the energy region we studied here. Numerical calculation using circuit design parameters (i.e., $l_0 = 0.43~{\rm nH/mm}$, $c_0 = 0.16~{\rm pF/mm}$, $L_j \approx 0.94~{\rm nH}$, $C_j \approx 11~{\rm fF}$, $l = 4.7~{\rm mm}$) shows that $g_1 / 2\pi \approx 2.53~{\rm GHz}$ and $g_2 / 2\pi \approx 2.70~{\rm GHz}$ near the optimal point ($\Phi_\mathrm{ext} / \varphi_0 \simeq \pi$), which qualitatively agree with the fitted values in the main text. A study of the influence of the fourth mode is presented in the next section.

\vspace{.4cm}
\noindent {\bf Influence of higher modes}\\
Although the physics of this work involves only the first two modes of the waveguide resonator, here we study the influence of the higher energy modes. The position of the qubit coincides with the node of the third mode, corresponding to a negligible coupling. Hence, we now study the influence of the fourth mode on both the eigenvalues and the transmission spectrum.

Supplementary Fig.~\ref{fig:fourth-mode-study}a shows the first eigenvalues of the total Hamiltonian with and without the fourth mode (solid blue and dashed orange lines, respectively), together with the bare qubit frequency (dash-dotted green line). Although the qubit is in resonance with the fourth mode at approximately $\delta\Phi_\mathrm{ext} \simeq 40 ~ {\rm m}\Phi_0$, the eigenvalues in the low energy region remain unperturbed. In Supplementary Fig.~\ref{fig:fourth-mode-study}b,c we show that also the transmission spectra are unaltered by the presence of the fourth mode. Again, the spectra is obtained using a low power input drive, involving less then one photon. This keeps the dynamics inside the low-energy subspace.

\begin{figure}[hbt!]
    \centering
    \includegraphics[width=\linewidth]{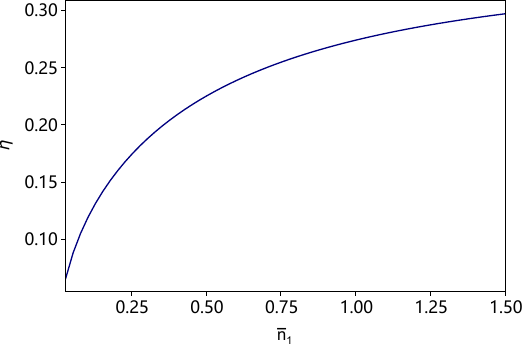}
    \caption{{Theoretically calculated} efficiency of the second harmonic generation as a function of the theoretical input photon number $\bar{n}_1$. This quantity is related to the number of photons emitted at frequency $2 \omega$ per unit input photon at $\omega$. This simulation was performed at $\delta\Phi_\mathrm{ext} = -45~{\rm m}\Phi_0$.}
    \label{fig:shg-ratio}
\end{figure}

\vspace{.4cm}
\noindent {\bf Efficiency of the Second Harmonic Generation}\\
Here we define the efficiency 
\begin{equation}
\eta \equiv S_{21}^{(2\omega)} / S_{21}^{(\omega)}
\end{equation} 
of the SHG as the ratio between the signal transmitted at $2 \omega$ over the one at $\omega$, when driving at $\omega$. This quantity gives us more information on the number of photons up-converted per unit incoming photon. Supplementary Fig.~\ref{fig:shg-ratio} shows the SHG efficiency as a function of the theoretical input photon number $\bar{n}_1$. This value is significant already for very small number of input photons.

%========================================================================
%========================================================================

\vspace{.4cm}
\noindent {\bf Input--output and generalized master equation}\\
%The standard quantum optics approach to describe open quantum systems no longer holds in the case of ultrastrong coupling. To properly derive the time evolution of the system and the input--output relations, the interaction term between the system and the environment has to be expressed on the basis of the eigenstates of the system. Moreover, the field input--output relations include anharmonic operators. This problem can be circumvented by expressing the time evolution of the field operators in terms of the generalized master equation, rather than the nonlinear input--output relations.
We start by considering the losses of the two-mode resonator field due to the presence of both the input and output ports, where the resonator is capacitively coupled to open coplanar waveguides. On the input side, the interaction term can be described as
\begin{equation}
    {\mathcal{V}}_l = i \int_0^{\infty} \dd{\omega} g_l (\omega) \left( {a}_{l, \omega} - {a}_{l,\omega}^\dagger \right) {X} \, ,
\end{equation}
where $g_l (\omega) = \sqrt{\kappa_{l} \omega / \omega_1}$ is the frequency dependent bath--resonator coupling rate (expressed in terms of the bare loss rate $\kappa_{l}$), and
\begin{equation}
    \label{eq: X operator input-output}
    {X} = i ({a}_1 - {a}_1^\dagger) + i \sqrt{\frac{\omega_2}{\omega_1}} ({a}_2 - {a}_2^\dagger) \, 
\end{equation}
is the adimensional two-mode field momentum.
Assuming the driving tone at frequency $\omega_\mathrm{in}$ to be in a coherent state, the input creation operator $\ket{\alpha_{\mathrm{in}}}$ can be decomposed into the sum of a  zero-mean quantum operator $\tilde{a}_{l, \omega}$ plus a $c$-number: ${a}_{l, \omega} = \tilde{a}_{\mathrm{l}, \omega} +  \delta (\omega - \omega_\mathrm{in}) \alpha_\mathrm{in}{\exp} (i \omega_\mathrm{in} t)$. Using $\alpha_\mathrm{in} = |\alpha_\mathrm{in}|\exp{i \phi}$, the resulting driving term is
\begin{equation}\label{Vin}
    {\mathcal{V}}_\mathrm{in} (t) = 2 |\alpha_\mathrm{in}| \sqrt{\kappa_{l}} \sqrt{\frac{\omega_\mathrm{in}}{\omega_1}} \sin (\omega_\mathrm{in} t + \phi) {X} \, .
\end{equation}
Here $|\alpha_\mathrm{in}|^2$ describes the rate of input photons and can be directly related to the input power: $P_{\rm in} = \hbar \omega_{\rm in} |\alpha_\mathrm{in}|^2$. 
Supplementary Eq.~\eqref{Vin} can be easily generalized to two (or more) driving tones. When considering a single tone, $\alpha_\mathrm{in}$ can be assumed to be real without any loss of generality. However, in the multi-tone case, the relative phase between different tones becomes relevant (see, e.g., Fig.~4 in the main text). On the output side, we have an analogous term
\begin{equation}
    \label{eq: interaction term output port}
    {\mathcal{V}}_{r} = i \int_0^{\infty} \dd{\omega} g_{r} (\omega) \left( {a}_{\mathrm{r}, \omega} - {a}_{\mathrm{r},\omega}^\dagger \right) {X} \, ,
\end{equation}
with $g_{r} (\omega) = \sqrt{\kappa_{{r}} \omega / \omega_1}$. In the Heisenberg picture, we define the positive frequency part of the output field operator~\cite{Milburn2008Springer} as
\begin{equation}
    {\Phi}_\mathrm{r,out}^+ (t) = \frac{1}{\sqrt{2 \pi}} \int_{0}^{\infty} \dd{\omega} \frac{1}{\sqrt{\omega}} e^{-i \omega (t - t_f)} {a}_{\mathrm{r}, \omega} (t_f) \, ,
\end{equation}
where $t_f$ is the time at which the measurement is performed, and we assume that there is no input field from the output port $\langle {\Phi}_\mathrm{r,in}^+ (t) \rangle = 0$. 

We now take the positive frequency part of the system field operator 
\begin{equation}
{X}^+ \equiv \sum_{j>k} {X}_{kj} 
\end{equation}
(${X}_{kj} \equiv X_{kj} \dyad{k}{j}$), which is written in the basis of the eigenstates of the system Hamiltonian, with $\ket{j}$ being the $j$-th eigenstate and $X_{kj} = \mel{k}{{X}}{j}$. By performing the rotating-wave approximation in Supplementary Eq.~\eqref{eq: interaction term output port}
\begin{equation}
    \left( {a}_{\mathrm{r}, \omega} - {a}_{\mathrm{r},\omega}^\dagger \right) {X} \simeq {a}_{\mathrm{r}, \omega} {X}^- - {a}_{\mathrm{r},\omega}^\dagger {X}^+ \, ,
\end{equation}
and following the standard input--output procedure~\cite{Milburn2008Springer}, the relation between ${\Phi}_\mathrm{out}^+$ and ${X}^+$ is
\begin{equation}
     \expval{{\Phi}_\mathrm{r,out}^+} (t) = \sqrt{\frac{\kappa_\mathrm{r}}{\omega_1}} \expval{{X}^+} (t) \, ,
\end{equation}
where the time dependence of $\langle {X}^+ \rangle (t)$ originates from the Hamiltonian dynamics, dissipations, and the coherent input field ${\mathcal{V}}_\mathrm{in} (t)$. Moreover, it is often useful to write the input--output relation in terms of $\dot{{\Phi}}_\mathrm{r, out}^+$ rather than ${\Phi}_\mathrm{r,out}^+$, because it is directly linked to the measurement of the voltage.

Compared to the standard treatment of driven-dissipative systems, the explicit time dependence cannot be traced out through a unitary transformation, due to the presence of the counter-rotating terms. This results in a time-dependent stroboscopic steady state ${\rho}_\mathrm{ss} (t)$ for $t \to \infty$, which, following the Floquet formalism, is periodic, with period $T = 2 \pi / \omega_\mathrm{in}$. Indeed, the time evolution of the total density matrix follows the generalized master equation $\dot{{\rho}} = \mathcal{L}_\mathrm{gme} (t) {\rho}$, where
\begin{equation}
    \label{eq: generalized me}
    \mathcal{L}_\mathrm{gme} (t) = \mathcal{L}_0 + \mathcal{L}_1 {\exp} (i \omega_\mathrm{in} t) + \mathcal{L}_{-1} {\exp} (-i \omega_\mathrm{in} t) \, .
\end{equation}
Here $\mathcal{L}_0$ describes the open evolution of the system in the absence of any external drive, taking into account losses of the resonator in both the input and output ports and possible additional internal loss. It also takes into account intrinsic qubit loss and pure dephasing with rates $\kappa_q$ and $\kappa_{q,\mathrm{dep}}$, respectively. Indeed, the Liouvillian can be written as~\cite{Settineri2018Dissipation}
\begin{equation}
    \label{eq: gme L_0}
    \begin{split}
        \mathcal{L}_0 {\rho} =& -i \comm{{H}_S}{{\rho}} + \frac{1}{2} 
        \sum_{\substack{n = (r_1,r_2,q) \\ j, k>j \\ l, m>l}} \tilde{\kappa}^{(n)} (\tilde{\omega}_{ml}) \times \\
        & \left\{ \; \left[ {X}_{lm}^{(n) \dagger} {\rho} {X}_{jk}^{(n)} - {X}_{jk}^{(n)} {X}_{lm}^{(n) \dagger} {\rho} \right] n_{\mathrm{th}} (\tilde{\omega}_{ml}, T_n) \right. \\
        & + \left[ {X}_{lm}^{(n)} {\rho} {X}_{jk}^{(n) \dagger} - {X}_{jk}^{(n) \dagger} {X}_{lm}^{(n)} {\rho} \right] \left[ n_{\mathrm{th}} (\tilde{\omega}_{ml}, T_n) + 1 \right] \\
        & + \left[ {X}_{jk}^{(n) \dagger} {\rho} {X}_{lm}^{(n)} - {\rho} {X}_{lm}^{(n)} {X}_{jk}^{(n) \dagger} \right] n_{\mathrm{th}} (\tilde{\omega}_{ml}, T_n) \\
        & + \left. \left[ {X}_{jk}^{(n)} {\rho} {X}_{lm}^{(n) \dagger} - {\rho} {X}_{lm}^{(n) \dagger} {X}_{jk}^{(n)} \right] \left[ n_{\mathrm{th}} (\tilde{\omega}_{ml}, T_n) + 1 \right] \right\} \\
        & + \kappa_{q,\mathrm{dep}} \mathcal{D} \bigg[ \sum_j {X}_{jj}^{(q)} \bigg] {\rho}\, ,
        % \left[ \vphantom{\sum} \right. \sum_j {X}_{jj}^{(q)} \left. \vphantom{\sum} \right] {\rho}
    \end{split}
\end{equation}
where
\begin{subequations}
    \begin{align}
        {X}^{(r_1)} = & \ i ({a}_1 - {a}_1^\dagger) + i \sqrt{\frac{\omega_2}{\omega_1}} ({a}_2 - {a}_2^\dagger) \, , \\\
        {X}^{(r_2)} = & \ \sqrt{\frac{\omega_2}{\omega_1}} ({a}_1 + {a}_1^\dagger) + ({a}_2 + {a}_2^\dagger) \, , \\
        {X}^{(q)} = & \ {\sigma}_z
    \end{align}
\end{subequations}
are the dissipation operators, and $\tilde \omega_{ml} = \tilde{\omega}_m - \tilde{\omega}_l$. Here ${X}^{(r_1)}$ is related to the dissipation induced by the interaction with the coplanar waveguide, ${X}^{(r_2)}$ is related to the internal loss through the inductance, and ${X}^{(q)}$ is related to the qubit dissipation. Moreover, 
\begin{subequations}
\begin{align}
\tilde{\kappa}^{(r_1)} (\omega) = & \  (\kappa_l + \kappa_r) \omega / \omega_1 \, ,\\\ 
\tilde{\kappa}^{(r_2)} (\omega) = & \  \kappa_\mathrm{int} \omega / \omega_1 \, ,\\
\tilde{\kappa}^{(q)} (\omega) = & \  \kappa_q \omega / \omega_1\, , 
\end{align}
\end{subequations}
and 
\begin{equation}
n_{\mathrm{th}} (\omega, T_n) = \left[ \exp (\hbar \omega / k_{\textrm{B}} T_n) - 1 \right]^{-1} 
\end{equation}
is the thermal population of the $n$-th reservoir describing the number of excitations at a given temperature $T_n$; $k_{\textrm{B}}$ is the Boltzmann constant, and
\begin{equation}
    \mathcal{D} \left[ {O} \right] {\rho} = \frac{1}{2} \left( 2 {O} {\rho} {O}^\dagger - {O}^\dagger {O} {\rho} - {\rho} {O}^\dagger {O} \right) 
\end{equation}
is the standard Lindblad dissipator. This term considers qubit pure dephasing effects due to stochastic fluctuations of the flux passing through the qubit, inducing additional decoherence effects.

It is worth to introduce real loss rates (directly related to the decay rate of a specific transition, corresponding to their spectral linewidths). For a given transition $\tilde{\omega}_{p,q}$ [which corresponds to the case where $(l,m) = (j,k) = (q,p)$ in Supplementary Eq.~\eqref{eq: gme L_0}], the total loss rate is then
\begin{equation}
    \Gamma_{p,q} = \tilde{\kappa}^{(r_1)}(\tilde{\omega}_{p,q}) \vert X_{pq}^{(r_1)} \vert^2 + \tilde{\kappa}^{(r_2)}(\tilde{\omega}_{p,q}) \vert X_{p,q}^{(r_2)} \vert^2 \, ,
\end{equation}
where we considered here only the resonator losses. Note that the relevant transitions studied here are not significantly affected by qubit decoherence.
As an example, the linewidth of the transition at $\tilde{\omega}_1$ is 
\begin{equation}
\Gamma_{1,0} = (\kappa_\mathrm{in} + \kappa_\mathrm{out}) \vert X_{10}^{(r_1)} \vert^2 + \kappa_\mathrm{int} \vert X_{10}^{(r_2)} \vert^2\, .
\end{equation}

The influence of the coherent input drive comes from $\mathcal{L}_\pm$, where
\begin{equation}
    \mathcal{L}_{\pm 1} {\rho} = \pm i \abs{\alpha_\mathrm{in}} e^{\pm i \phi} \sqrt{\kappa_\mathrm{in}} \sqrt{\frac{\omega_\mathrm{in}}{\omega_1}} \comm{{X}}{{\rho}} \, .
\end{equation}
Following Supplementary Eq.~\eqref{eq: generalized me}, we can expand the steady state in Fourier components ${\rho}_\mathrm{ss} (t) = \sum_{n=-\infty}^{+\infty} {\rho}_n {\exp} (i n \omega_\mathrm{in} t)$. By putting this form of the density matrix into Supplementary Eq.~\eqref{eq: generalized me}, and equating each Fourier component, we obtain a tridiagonal recursion relation
\begin{equation}
    \left( \mathcal{L}_0 - i n \omega_\mathrm{in} \right) {\rho}_n + \mathcal{L}_1 {\rho}_{n-1} + \mathcal{L}_{-1} {\rho}_{n+1} = 0 \, ,
\end{equation}
and impose that ${\rho}_m = 0$ for a sufficiently large $m$.

The coherent emission intensity of the resonator at a frequency $n \omega_\mathrm{in}$ is proportional to ${\Tr} [{X}^+ {\rho}_{-n}]$. For the transmission spectrum, where the coherent input  $|\alpha_\mathrm{in}| / \sqrt{\omega_\mathrm{in}}$ at frequency $\omega_\mathrm{in}$ is compared to the coherent output ${\Phi}_\mathrm{r,out}^+$ at the same frequency, we have
\begin{equation}
    S_{21} = \sqrt{\frac{\omega_\mathrm{in}}{\omega_1}}  \frac{\sqrt{\kappa_\mathrm{r}}}{\abs{\alpha_\mathrm{in}}} \Tr [{X}^+ {\rho}_{-1}] \, .
\end{equation}
In the simple case of a single harmonic oscillator without internal loss ($\kappa_\mathrm{int} = 0$), we have 
\begin{equation}
\Tr [{X}^+ {\rho}_{-1}] = 2 \abs{\alpha_\mathrm{in}} \sqrt{\kappa_{r}} / (\kappa_{r} + \kappa_{l}) 
\end{equation}
on resonance, and the standard transmission formula is obtained~\cite{Milburn2008Springer}.
%========================================================================
\begin{figure}[hbt!]
    \centering
    \includegraphics[width=\linewidth]{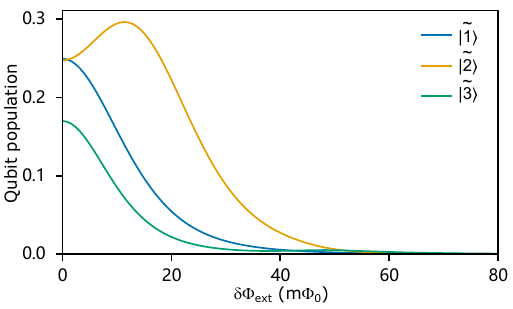}
    \caption{Population of the dressed qubit as a function of the flux bias $\delta \Phi_\mathrm{ext}$. The population is computed as $\langle {\Sigma}^- {\Sigma}^+ \rangle$, with ${\Sigma}^+ = \sum_{j>k} \langle k \vert {\sigma}_z \vert j \rangle \vert k \rangle \langle j \vert$, and it is almost negligible in the region around $\delta\Phi_\mathrm{ext} \simeq 47 ~ {\rm m}\Phi_0$, demonstrating the pure-photonic nature of the effect.}
    \label{fig:qubit-population}
\end{figure}
%========================================================================
For the SHG spectrum, we calculate the emission from the component oscillating at $2 \omega_\mathrm{in}$. In other words,
\begin{equation}
    S_\mathrm{SHG} \propto \Tr [ {X}^+ {\rho}_{-2} ] \, .
\end{equation}
In the case of the interference pattern in Fig.~4 of the main text, we used a drive with harmonics of frequency $\omega_\mathrm{in}^{(1)}$ and $\omega_\mathrm{in}^{(2)}$. In the first case, depicted in Fig.~4a,b of the main text, the interaction term ${\mathcal{V}}_\mathrm{in}$ on the left side becomes
\begin{equation}\label{Vin2}
    \begin{split}
        {\mathcal{V}}_\mathrm{in} (t) = & 2 \sqrt{\kappa_{l}} \left[ |\alpha_\mathrm{in}^{(1)}| \sqrt{\frac{\omega_\mathrm{in}^{(1)}}{\omega_1}} \sin (\omega_\mathrm{in}^{(1)} t) \right. \\
        & + \left. |\alpha_\mathrm{in}^{(2)}| \sqrt{\frac{\omega_\mathrm{in}^{(2)}}{\omega_1}} \sin (\omega_\mathrm{in}^{(2)} t + \phi) \right] {X} \, ,
    \end{split}
\end{equation}
where $\phi$ is the relative phase between the two tones, and we vary the intensity of the second tone at frequency $\omega_\mathrm{in}^{(2)} = 2 \omega_\mathrm{in}^{(1)}$. In the second case, as shown in Fig.~4c,d of the main text, we vary the intensity of the first tone, with the phase factor inside the first driving term rather than the second one. Since in the device presented here, the left port corresponds to the input port, while the right one is the output port, throughout this article we define $\kappa_{\rm in} \equiv \kappa_l$ and $\kappa_{\rm out} \equiv \kappa_r$ (see Supplementary Fig.~\ref{fig:S1}).
The estimation of the number of photons is performed following the standard formula of a simple driven-dissipative harmonic oscillator. Although in the USC regime, this might slightly change, it provides a simple formula relating the input power with the mean number of photons in the resonator and helps for comparison with standard quantum optics architectures. Thus, the mean number of photons in the first (second) mode is evaluated as
\begin{equation}
    \bar{n}_{1} = 4 \frac{P_\mathrm{in}^{1} \kappa_{\rm in}}{\omega_1 \Gamma_{1,0}^2} \, , \qquad \qquad \bar{n}_{2} = 4 \frac{P_\mathrm{in}^{2} \kappa_{\rm in}}{\omega_2 \Gamma_{3,0}^2} \, .
\end{equation}

%========================================================================
\begin{figure}[hbt!]
    \centering
    \includegraphics[width=\linewidth]{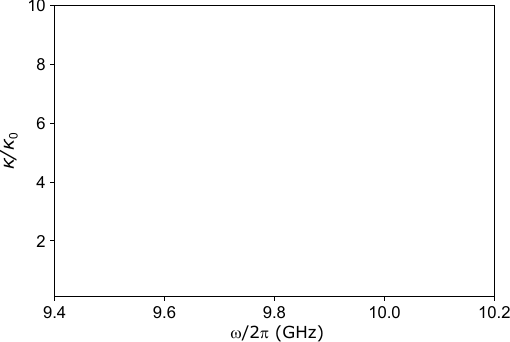}
    \caption{Rabi splitting visibility as a function of the qubit losses $\kappa_q$ and $\kappa_{q,\mathrm{dep}}$ for both internal losses and pure dephasing. The value of the losses is normalized by the ones used in this work. As can be seen, aside from a small attenuation of the intensity, the Rabi splitting remains visible and distinguishable even at large qubit losses.}
    \label{fig:spectrum-vs-kappa}
\end{figure}
%========================================================================

\vspace{.4cm}
\noindent {\bf Phase sensitivity of parametric amplification}\\
The evolution equation for a linear degenerate parametric amplifier is (see, e.g., \cite{Caves1982})
\begin{equation}
     b =  a\, {\rm cosh}r + e^{i \phi} a\, {\rm sinh}r\, ,
\end{equation}
where $ a$ and $ b$ are the input and output destruction photon operators at $\omega$,
$r$ is a real constant determined by the strength and duration of the interaction (proportional to the pump amplitude at $2 \omega$, and $\phi$ describes the pump phase. Assuming a coherent input at $\omega$, $ a \to \alpha$ with $\alpha$ real (corresponding to a relative phase between the two tones $\phi$), in the weak $r$ limit, we obtain
\begin{equation}
   \langle  b \rangle = \alpha (1 +e^{i\phi} r)\, 
\end{equation}
and thus
\begin{equation}
   |\langle  b \rangle|^2 = \alpha^2 (1 + r \cos \phi)^2 + r^2 \sin^2 \phi\, ,
\end{equation}
showing an output signal at $\omega$ depending on both the phase and amplitude of the drive at $2 \omega$, in qualitative agreement with the data shown in Fig. 4a in the main text.

\vspace{.4cm}
\noindent {\bf Dependence of the Rabi splitting on the qubit decoherence rate}\\
Here we demonstrate that, despite the non-negligible qubit decoherence in our device, the visibility of the effect (e.g., the Rabi splitting in Fig.~2 of the main text) is not significantly impacted by the qubit decoherence. Indeed, the qubit is very far detuned from the photonic resonance frequencies, especially in the region of the avoided-level-crossing. For this reason{, as also shown in Supplementary Fig.~\ref{fig:qubit-population},} its participation in the dynamics is minimal. Supplementary Fig.~\ref{fig:spectrum-vs-kappa} shows the photonic Rabi splitting as in Fig.~2 of the main text, as a function of the qubit losses $\kappa_q$ and $\kappa_{q,\mathrm{dep}}$ for both internal losses and pure dephasing. The value of the losses is normalized by the ones used throughout the work. As already explained, the qubit participation is very small, and it does not affect the visibility, aside from a small attenuation. Indeed, we observe no significant linewidth broadening as the qubit losses increase. This fact is very important, because many qubits are susceptible to noise. Thus, this shows the robustness of our setup under nearly all conditions.

%========================================================================
\begin{figure}[t]
    \centering
    \includegraphics[width=\linewidth]{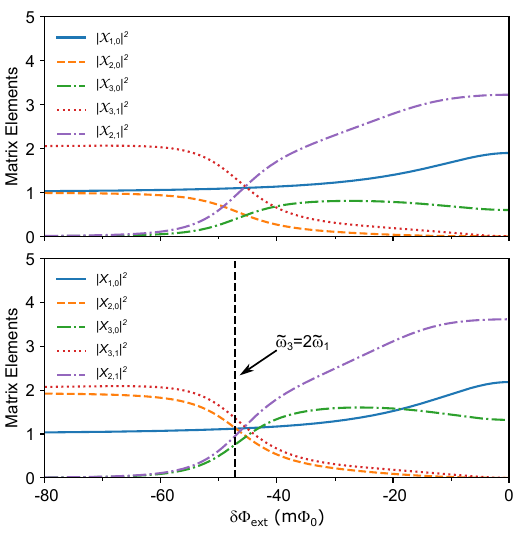}
    \caption{Relevant photonic transition matrix elements between the lowest energy levels of the system as a function of the flux offset. The matrix elements are calculated for both the input--output dissipation operator ${X}$ (bottom panel) and ${\mathcal{X}}$ (upper panel).}
    \label{fig:S6}
\end{figure}
%========================================================================

%========================================================================
\begin{figure}[t]
    \centering
    \includegraphics[width=\linewidth]{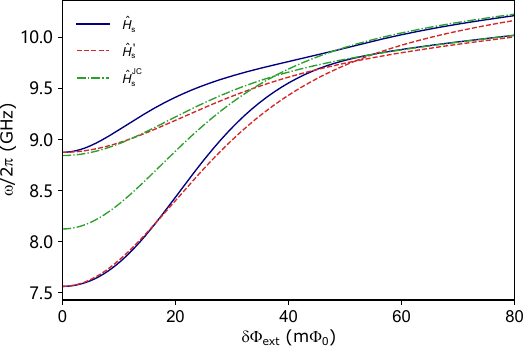}
    \caption{Energy levels in the region of the one-two-photon avoided level crossing obtained by diagonalizing three different Hamiltonians. The case with ${H_\mathrm{s}}$ is the only one showing the avoided level crossing, proving that both the counter-rotating terms and the parity symmetry breaking are needed to achieve this effect.}
    \label{fig:S7}
\end{figure}
%========================================================================

\vspace{.4cm}
\noindent {\bf Photonic transition matrix elements}\\
Many peculiar behaviors of the circuit-QED system investigated here can be understood based on a few photonic transition matrix elements, describing the strength of transitions between the lowest energy levels of the system. We calculated them by direct numerical diagonalization of the Hamiltonian in Eq.~(1) of the main text, after fixing parameters by fitting the experimental spectral lines in Fig.~2 of the main text. 

Supplementary Fig.~\ref{fig:S6} shows the calculated off-diagonal matrix elements $\langle E_j|{\cal X} |E_k \rangle$ and $\langle E_j|{X} |E_k \rangle$, as a function of the flux offset, where $ { {\cal X}} = {a}_1 + {a}_1^\dagger + {a}_2 +{a}_2^\dagger$ and ${X}$ as in Supplementary Eq.~\eqref{eq: X operator input-output}. The eigenstates $|E_j\rangle$ of the Hamiltonian in Eq.~(1) of the main text are sorted in ascending order so that $j > k$ for energy eigenvalues $\tilde \omega_j  > \tilde \omega_k$. State $|E_1\rangle \equiv |\widetilde{1,0}\rangle \simeq |1,0,g\rangle$ corresponds to the dressed one-photon (of mode $n=1$) state. The impact of the hybridization of one- and two-photon states around the avoided-level crossing ($\delta \Phi_{\rm ext} \approx -47 ~\rm m \Phi_0$) is evident.
For small absolute values of the flux offset $|\delta \Phi_{\rm ext}|$, $|E_2\rangle \simeq |2,0,g\rangle$, and $|E_3\rangle \simeq |0,1,g\rangle$, while in the opposite limit, after the avoided-level crossing, ($|\delta \Phi_{\rm ext}| > 80~\rm m \Phi_0$), $|E_2 \rangle \simeq |0,1,g\rangle$, and $|E_3\rangle \simeq |2,0,g\rangle$. The vertical dashed line indicates the degeneracy condition where SHG occurs. Note that at this flux offset ($\delta \Phi_{\rm ext} \sim -47~\rm m\Phi_0$) $|{X}_{1,0}|^2 \sim 1.2$, $|{X}_{3,1}|^2 \sim 1.4$, while $|{X}_{3,0}|^2 \sim 0.8$. The one--two-photon strong coupling gives rise to a harmonic three-level $\Delta$ system ($|E_0\rangle$, $|E_1\rangle$, $|E_3\rangle$), where $\tilde \omega_3 - \tilde \omega_1 = \tilde \omega_1 - \tilde \omega_0$, with all transition matrix elements comparable and close to the standard matrix element $\langle 0| a | 1 \rangle = 1$ for one-photon transition in vacuum. The matrix elements in Supplementary Fig.~\ref{fig:S6} enter in all the theoretical calculation displayed in this article and determine the observed nonlinear optical processes below the single-photon power level. Note that at $\delta \Phi_{\rm ext} \sim -47~\rm m\Phi_0$ state $|E_3 \rangle$ corresponds to the upper energy state $|\psi_+ \rangle$ in the one--two-photon avoided level crossing, while $|E_2 \rangle \equiv |\psi_- \rangle$ to the lower energy state.

\vspace{.4cm}
\noindent {\bf Comparison with other Hamiltonians}\\
It is useful to compare our model with other two cases. The first one without considering the parity symmetry breaking term, and the second one without including the counter-rotating terms. First, we express the original Hamiltonian in Eq.~(1) of the main text using the basis where the qubit Hamiltonian is diagonal:
\begin{equation}
    \begin{split}
        {H}_\mathrm{s} = & \frac{\omega_q}{2} {\sigma}_z + \sum_{n=1,2} \left[\omega_n {a}_n^{\dagger}{a}_n \right. \\
        & + \left. {g}_n({a}_n^{\dagger}+{a}_n) (-\sin (\theta) {\sigma}_x + \cos(\theta) {\sigma}_z )
\right].
    \end{split}
\end{equation}
Now, we can define the Hamiltonian ${H}_\mathrm{s}^\prime$, obtained by neglecting the parity-symmetry breaking term in the interaction part:
\begin{equation}
        {H}_\mathrm{s}^\prime = \frac{\omega_q}{2} {\sigma}_z + \sum_{n=1,2} \left[\omega_n {a}_n^{\dagger}{a}_n \right. - \left. \sin(\theta) {g}_n({a}_n^{\dagger}+{a}_n) {\sigma}_x\right],
\end{equation}
and the Jaynes--Cummings equivalent Hamiltonian
\begin{equation}
        {H}_\mathrm{s}^{\mathrm{JC}} = \frac{\omega_q}{2} {\sigma}_z + \sum_{n=1,2} \left[\omega_n {a}_n^{\dagger}{a}_n \right. - \left. \sin(\theta) {g}_n({a}_n^{\dagger} {\sigma}_- + {a}_n {\sigma}_+)\right].
\end{equation}
Supplementary Fig.~\ref{fig:S7} shows the energy levels of these three Hamiltonians, proving that both the parity symmetry breaking and the counter-rotating terms are needed to achieve the avoided-level crossing (and thus the strong coupling between one and two photons).

%%%%%%%%%%%%%%%%%%%%%%% References %%%%%%%%%%%%%%%%%%%%%%%%%	